\documentclass[twocolumn,showpacs,preprintnumbers,amsmath,amssymb,prb,aps]{revtex4-1}
\usepackage{epsfig}% Include figure files
\usepackage{epstopdf}
\usepackage{graphicx}% Include figure files
\usepackage{dcolumn}% Align table columns on decimal point
\usepackage{bm}% bold math
\usepackage{textcomp}
\usepackage{array}
\usepackage{subcaption}
\usepackage{booktabs}

\begin{document}

\title{A detailed electronic structure study of Vanadium metal by using different beyond-DFT methods}
\author{Antik Sihi$^{1,}$}
\altaffiliation{sihiantik10@gmail.com}
\author{Sudhir K. Pandey$^{2,}$}
\altaffiliation{sudhir@iitmandi.ac.in}
\affiliation{$^{1}$School of Basic Science, Indian Institute of Technology Mandi, Kamand - 175005, India\\
$^{2}$School of Engineering, Indian Institute of Technology Mandi, Kamand - 175005, India}

\date{\today}

\begin{abstract}

 We report a detailed electronic structure calculation for Vanadium (V) using DFT, DFT+$U$, $G_0W_0$, $GW_0$ and DFT+DMFT methods. The calculated values of $W$, $U$ and $J$ by cRPA method are $\sim$1.1, $\sim$3.4 and $\sim$0.52 eV, respectively. The comparison between calculated spectra (CS) and experimental spectra (ES) suggests that $W$ ($U$) is more accurate for DFT+$U$ (DFT+DMFT) method. The CS, obtained by these methods, give fairly good agreement with ES for peaks' positions except $GW_0$. The shallowness of the dips lying $\sim$ -1.5 eV and $\sim$1.0 eV in ES are properly explained by DFT+DMFT method only, due to the presence of incoherent $t_{2g}$ states. This work suggests that for the proper explanation of ES, sophisticated many-body theory is needed even for the simple metal. \\

\end{abstract}

\maketitle

\section{Introduction} 

 Generally, any material containing partially filled $d$ and/or $f$ -block elements of the periodic table is known as the strongly correlated electron systems (SCES). In recent years, the ongoing research on these systems have opened a new era of research in condensed-matter physics for their emergent physical properties \cite{kotliar,edelstein}. Therefore, depending on different physical properties, these materials are subdivided into different systems for example heavy fermion \cite{stewart}, heavy fermion superconductors \cite{ott}, Kondo insulators \cite{fisk}, non-Fermi liquid systems \cite{aronson} etc. However, the effect of strong electron-electron correlations are carrying very important role for these materials to obtain different kind of new physical properties. Theoretically, to understand the depth of correlation effect for any material, one important parameter is needed to be calculated which is known as the on-site Coulomb interaction ($U$). After the recent theoretical development, it is possible to calculate $U$ by two different ways, which are known as constrained density functional theory (cDFT) \cite{anisimov} and constrained random-phase approximation (cRPA) \cite{aryasetiawan,vaugier}. It is known that cDFT gives larger value of $U$ than cRPA for late transition metals due to consideration of self-screening effect of the localized orbitals \cite{karlsson,miyake}. But, to understand the physical properties of the materials, both static and frequency-dependent $U$ are possible to calculate by using cRPA method with excluding the suitable choice of self-screening in localized orbitals. The fundamental way to verify the validation of calculated $U$ is done by comparing the density of states (DOS) with experimental spectra (ES) as obtained from different spectroscopic techniques like x-ray-photoemission spectroscopy (XPS), bremsstrahlung isochromat spectroscopy (BIS), inverse photoemission spectroscopy etc. \cite{paromita1,paromita2} 
 
 Theoretically, it is really difficult to find out the solution of many-electron problem. Nowadays, two different approaches are mainly implemented for tackling this problem viz. (I) model Hamiltonian based for low lying energy around the Fermi level (E$_F$) using some parameters and (II) first-principle based methods. The well known and successful first-principle based calculation is density functional theory (DFT). In this theory, the many-electron problem is solved with the help of independent-electron picture which provides reasonably good agreement with experimental data for various physical properties of many materials \cite{payne,shastri}. However, it is known that DFT fails to give the proper information about the electronic and magnetic properties for SCES \cite{terakura,sawatzky}. Also, the information about excited-state spectra for any material are difficult to obtain by using DFT method only \cite{georges,savrasov}.
 
  The reason for failure of DFT in case of SCES is that it does not consider the localization effect of $d$/$f$ orbitals properly. Although, the most common way to solve this problem is adding on-site interaction $U$ in DFT, which is known as DFT+$U$ method \cite{anisimov1}. This technique provides quite good result to predict the insulating ground state for many materials as observed from experiment, where DFT has shown the metallic ground state for same materials \cite{anisimov1}. But, some of the spectroscopic observation for $f$-electron system is not properly explained by DFT+$U$ method \cite{zhu,arko,havela}. It is also found that in DFT+$U$ method, $U$ is used as a parameter. So, for understanding the different physical properties of one material, different values of $U$ are used for the same material. This is an unphysical situation for tailoring or theoretical prediction of new materials. Therefore, the need of parameter free electronic structure calculation is very necessary to improve the strength of theoretical prediction. 
  
  In present days, the $GW$ approximation (GWA) based on many-body perturbation theory is known for a parameter free electronic structure method as developed by L. Hedin \cite{hedin}. It is basically the Hartree-Fock approximation including dynamical screening effect of Coulomb interaction \cite{gunnersson}. Therefore, the $GW$ self-energy ($\Sigma^{GW}$) depends both on crystal momentum \textbf{k} and frequency ($\omega$). However, the most well established all electron $GW$ based technique is known as one-shot $GW$ ($G_0W_0$) due to it's low computational cost. In case of band gap, this technique shows a fairly good matching with the experimental data than DFT method \cite{lebegue,hybersten}. But, the most serious problems with the $G_0W_0$ method is that this technique does not satisfy the conservation laws of momentum, energy and particle number \cite{baym,kadanoff,dahlen}. While, these problems are partially solved by conserving the particle number in $GW_0$ method \cite{martin}. In this method, the Dyson equation is solved by fully self-consistent one particle Green's function ($G$) with a fixed $W_0$ \cite{martin}. It is known that the spectral function and band gaps obtained from $G_0W_0$ are nicely agreed with the experimental data for those materials which contain open $s$ and/or $p$ -block elements of periodic table \cite{martin}. Thus in general, the GWA shows quite good agreement with experimental value for weakly correlated materials. But in materials, when the on-site Coulomb interaction has strong frequency dependence, the GWA is inadequate for the proper explanation of different physical properties of these materials \cite{nilsson,adler,nilsson2}.   
 
 One of the most advanced theoretical method to study many-electron problems as well as the strong correlation effect is known as dynamical mean field theory (DMFT) on top of DFT calculation ($i.e.$ DFT+DMFT). This method is capable to explain many experimental observation of the SCES \cite{kotliar,haule}. It is also noted that this technique has successfully described the localization effect and quasiparticle excitation at finite temperature. It deals with a well known localized impurity problem, where the self-energy ($\Sigma$) is the function of $\omega$ only. This $\Sigma (\omega)$ has important contribution for obtaining the quasiparticle excitation \cite{martin}. Further, it is also found that the different physical properties of many 3$d$ transition metals are nicely explained by using DFT+DMFT\cite{grechnev,chadov}.
 
 For long time, 3$d$ transition metals have drawn much attention of the researcher to understand their many emergent physical properties. From those, one of the most important 3$d$ transition metal is Vanadium (V) in body-centered cubic (BCC) crystal structure \cite{takemura,aschroft,wijn}. This metal is a Pauli paramagnet \cite{wijn,blundell}. It is also a well known type II superconductor with critical transition temperature ($T_c$) at 5.3 K, where the transition from superconducting state to normal metallic state is observed \cite{wexler,louis}. Although, both experimentally and theoretically, it is observed that $T_c$ for V can be tuned up to 17.2 K with increasing the pressure up to 1.2 Mbar \cite{louis,ishizuka}. Additionally, it shows a structural phase transition from BCC to simple cubic on applying the pressure $\sim$1.37 Mbar \cite{louis,suzuki}. 
 
 In this work, we focus to find the suitability of above mentioned methods for understanding the spectral properties of paramagnetic electron system in simple structure. So, V is the right candidate to choose for this study. Here, $U$ and on-site exchange interaction ($J$) for this metal are calculated by using cRPA method. The significance of $\omega$ dependency on $U$ and $W$ are explained for V. From this explanation, it is also justified that this metal is belonging to the class of correlated electron system. In order to benchmark the suitable electronic structure method for this class of system, different $ab$ $initio$ techniques ($e.g.$ DFT, DFT+$U$, $G_0W_0$, $GW_0$ and DFT+DMFT) are used in this work. The calculated $W$ ($U$) using cRPA method is used for DFT+$U$ (DFT+DMFT) calculation. For peaks' positions, all methods give good agreement with ES except $GW_0$. But, the proper shallowness of the dips as shown in ES are only obtained by DFT+DMFT (at 300 K) method. The momentum-resolved spectral function is plotted for comparing with the DFT band structure and to discuss about the coherent and incoherent states of V. The effect of $\Sigma (\omega)$ on the spectral function of V is discussed and the calculated value of effective band mass-renormalization parameter ($m^*$) is 1.14 (1.22) for $e_g$ ($t_{2g}$) orbitals.

\section{Computational details}

 Here, the spin-unpolarized electronic structure calculation for V is carried out by using the full-potential linearized-augmented plane-wave method. The local density approximation (LDA) is used as exchange-correlation functional throughout this calculation. The space group of \textit{I}m-3m and the lattice parameter of 3.024 \AA\ are used for this calculation \cite{mincryst}. 10 $\times$ 10 $\times$ 10 k-mesh size is used with fixed the convergence criteria at 10$^{-4}$ Ry/cell for total energy. DFT calculation is performed by using WIEN2k code \cite{blaha}. In order to calculate the Hubbard $U$, the cRPA method with Wannier basis function is used, which is implemented in GAP2 code\cite{jiang,li}. This code is interfaced with WIEN2k \cite{blaha}. Here, the density-density type of interactions is considered as Hubbard $U$. The GAP2 is also used for $G_0W_0$ and $GW_0$ calculations. DFT+$U$ calculation is done by Elk code \cite{elk} because its' implementation is the most general and simple \cite{bultmark}.

  DFT+DMFT calculation at 300 K is carried out using eDMFT code \cite{yee}, which is interfaced with WIEN2k \cite{blaha}. Dense 36 $\times$ 36 $\times$ 36 k-mesh size is used for this calculation as $\Sigma(\omega)$ and the features of total density of states (TDOS) are found to be sensitive for lower k-mesh. The continuous-time quantum Monte Carlo method is used as an impurity solver \cite{khaule}. To get rid from the double-counting problem, \textquoteleft exactd\textquoteright \, method is used \cite{hauleprl}. To plot the spectral function in real axis, maximum entropy analytical continuation method is used \cite{jarrell}.     

\section{Results and Discussion} 

  The dispersion curve of V using DFT along the high symmetric k-directions is shown in Fig. 1. In case of V, it is known that 4$s$ state has lower energy than 3$d$ state. From the figure, bands 2 to 6 are mainly contributed by 3$d$ orbitals, whereas in band 1, 4$s$ orbital also contributes. Hence, 1 to 6 energy bands are chosen to exclude the electronic transitions of 3$d$-3$d$ state for finding the value of $U$ and $J$ using cRPA method. It is noted that for $e_g$ ($t_{2g}$) orbitals, the computed values of intra bare Coulomb interaction, intra fully screened Coulomb interaction and intra Coulomb interaction ($U_{intra}$) are $\sim$18.67 ($\sim$17.93) eV, $\sim$1.30 ($\sim$1.01) eV and $\sim$3.58 ($\sim$3.24) eV, respectively. Moreover, the matrix elements of inter Coulomb interaction ($U_{inter}$) and $J$ for $e_g - e_g$ ($t_{2g} - t_{2g}$) orbitals are found to be $\sim$2.27 ($\sim$2.17) eV and $\sim$0.65 ($\sim$0.53) eV, respectively. Whereas, the calculated values of $U_{inter}$ ($J$) for $d_{z^2} - d_{xy}$, $d_{z^2} - d_{xz}$, $d_{z^2} - d_{yz}$, $d_{x^2-y^2} - d_{xy}$, $d_{x^2-y^2} - d_{xz}$ and $d_{x^2-y^2} - d_{yz}$ are $\sim$2.15 ($\sim$0.63) eV, $\sim$2.57 ($\sim$0.42) eV, $\sim$2.57 ($\sim$0.42) eV, $\sim$2.71 ($\sim$0.34) eV, $\sim$2.29 ($\sim$0.56) eV and $\sim$2.29 ($\sim$0.56) eV, respectively. Here, the full Coulomb interaction ($U_{full}$) (diagonal Coulomb interaction ($U_{diag}$)) denotes the value which is obtained by averaging the all (diagonal) elements of the Coulomb interaction matrix. The calculated values of $U_{full}$, $U_{diag}$ and $J$ are $\sim$2.6 eV, $\sim$3.4 eV and $\sim$0.52 eV, respectively. The unscreened (bare) Coulomb interaction and fully screened Coulomb interaction ($W$) are also calculated. The diagonal bare Coulomb interaction, full bare Coulomb interaction, on-site bare exchange interaction and $W$ are found to be $\sim$17.23 eV, $\sim$18.23 eV, $\sim$0.62 eV and $\sim$1.1 eV, respectively.

\begin{figure}
  \begin{center}
    \includegraphics[width=0.75\linewidth, height=4.5cm]{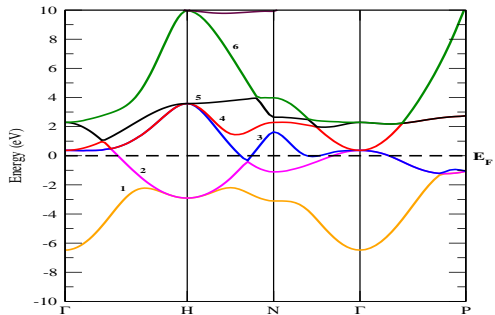} 
    \caption{(Colour online) Electronic band structure using DFT. Zero energy represents the Fermi level.}
    \label{fig:}
  \end{center}
\end{figure} 

\begin{figure}
  \begin{center}
    \includegraphics[width=0.75\linewidth, height=4.5cm]{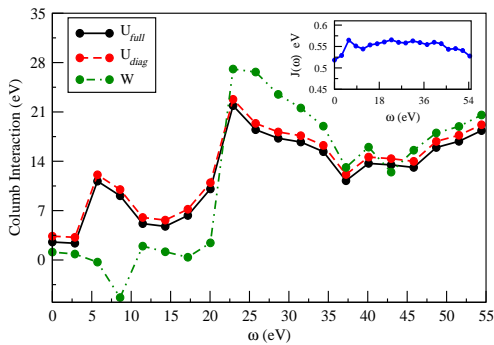} 
    \caption{(Colour online) Coulomb interaction with respect to $\omega$. $J (\omega)$ is shown in the inset.}
    \label{fig:}
  \end{center}
\end{figure}

\begin{figure*}
  \begin{center}
    \includegraphics[width=0.9\linewidth, height=14cm]{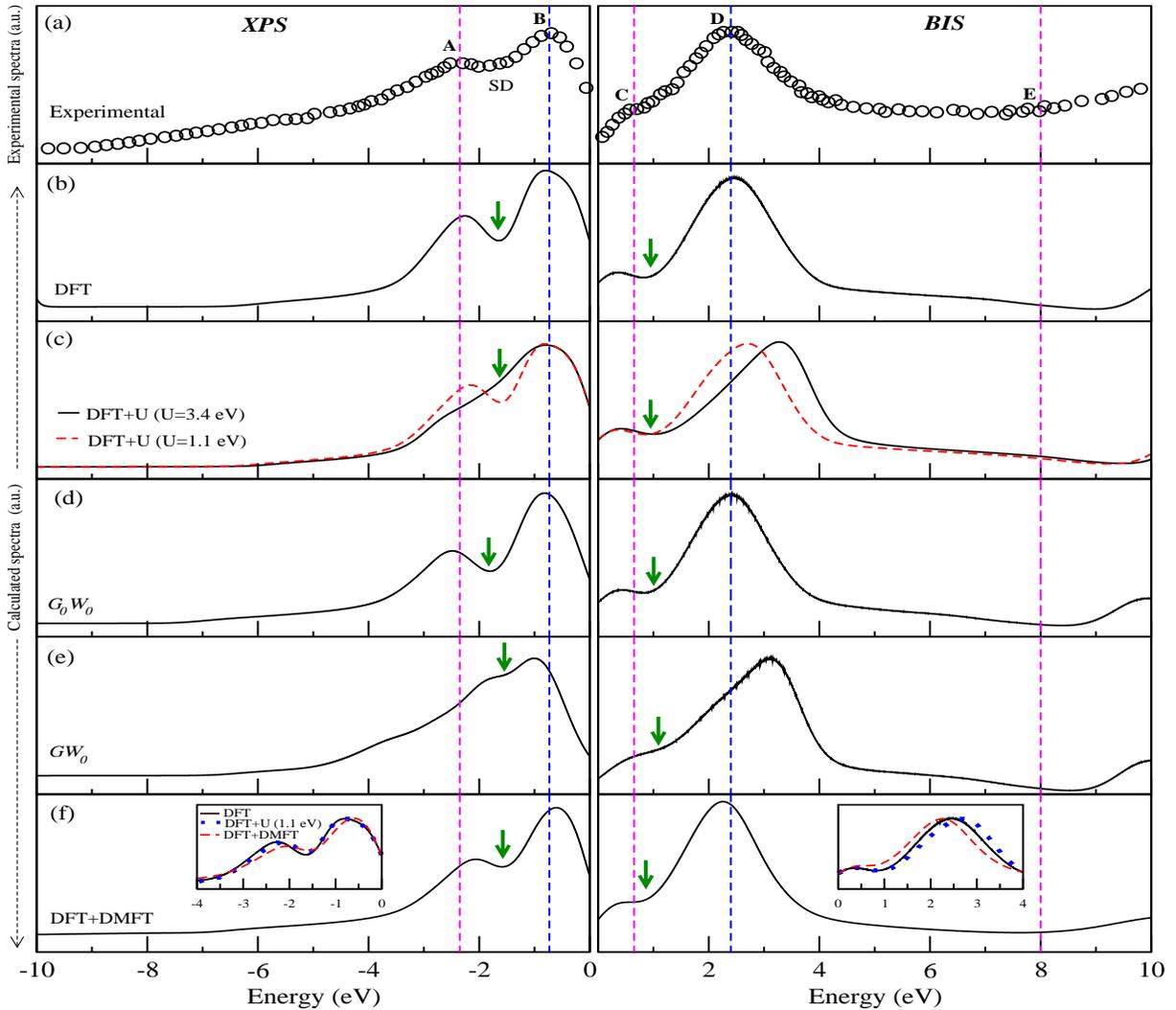} 
    \caption{(Colour online) (a) Experimental valence band using XPS \cite{ley} and conduction band using BIS \cite{speier}, calculated spectra using (b) DFT, (c) DFT+$U$, (d) $G_0W_0$, (e) $GW_0$ and (f) DFT+DMFT (at 300 K). Zero energy represents the Fermi level.}
    \label{fig:}
  \end{center}
\end{figure*}

  Now, the importance of orbital screening effect for V is discussed with the help of $\omega$ dependent $U_{full}$, $U_{diag}$ and $W$ plots, as shown in Fig. 2. It is clearly observed from the figure that in higher $\omega$ region, $U_{full}$, $U_{diag}$ and $W$ are close to bare Coulomb interaction due to negligible screening effect. But, as the $\omega$ goes below $\sim$22 eV, the values of $U_{full}$, $U_{diag}$ and $W$ have drastically decreased, which implies that screening is more effective in lower $\omega$ region. It is also noted from the figure that $U_{full}$ and $U_{diag}$ are significantly larger than $W$ in low frequencies. This behaviour represents that the elimination of 3$d$-3$d$ transitions is very important for finding the suitable material specific $U$. Moreover based on this result, it is expected that only adding a static $U$ in any theoretical calculation may not provide a good explanation of experimental data for V. However, consideration of $\omega$ dependence $U$ $i.e.$ $U(\omega)$ within DFT+DMFT may provide reasonable explanation of the experimental data. Further, the exchange interaction $J (\omega)$ is shown in the inset of Fig. 2, which remains almost the same for all values of $\omega$ studied here. For further calculations the values of $U$, $W$ and $J$ at $\omega$=0 are used. At this point, it is important to note that the Wannier basis function is used for calculating the $U$, $W$ and $J$, whereas different basis function is used for the calculation of DFT+$U$ \cite{bultmark} and DFT+DMFT \cite{yee}. The calculated values of $U$, $W$ and $J$ normally show the dependency on the used basis functions which is small as shown in the work of Miyake $et$ $al$ \cite{miyake}. Such a small variation in $U$, $W$ and $J$ are not expected to change the final conclusion of the present work.
 
  In this work, $U_{diag}$ is used for further calculations and represented as $U$ (= 3.4 eV), which is in good agreement with other previous work \cite{imada,sasioglu}. The validation of any electronic structure calculation is normally done by comparing the experimental data of different spectroscopic techniques such as XPS for valence band (VB) DOS and BIS for conduction band (CB) DOS. But one should keep in mind that these ES contained the information about partial density of states (PDOS) with their photoionization cross-sections, lifetime broadening of individual states, inelastic scattering background and instrumental broadening. However, it is known that the shape of lifetime (instrumental)  broadening is Lorentzian (Gaussian) type. Here, to compare with ES, the theoretical spectra are calculated. In order to calculate this spectra, the TDOS is chosen, which is multiplied by Fermi-Dirac (FD) distribution (subtracted amount of FD distribution from unity) for VB (CB) at 300 K with consideration of E$_F$ at zero. Then, these TDOS are convoluted with a constant Lorentzian broadening of 0.1 eV both for VB and CB, along with a Gaussian broadening of 0.55 eV (0.7 eV) is used for VB (CB) corresponding to experimental resolution. The photoionization cross-sections of 3$d$ and 4$s$ orbitals are of the same order in magnitude \cite{lindau}. Thus, TDOS is chosen for CS. The ES and all CS using different theoretical techniques are shown in Fig. 3(a) to 3(f) within the energy -10 eV to 10 eV.      
  
  The ES of V are shown in Fig. 3(a) for VB (CB), which are obtained using XPS \cite{ley} (BIS \cite{speier}). The three experimental peaks marked as A, B and D are observed at $\sim$ -2.4 eV, $\sim$ -0.7 eV, and $\sim$2.4 eV, respectively. In CB, one hump is also seen at $\sim$0.7 eV, which is marked by C. Moreover, a monotonically increasing behaviour in CB is observed after 8.0 eV. In order to focus on this feature, a dashed line E is marked at energy 8.0 eV. It is clear from the figure that peaks A \& B are in VB and peak D is in CB. In VB, a dip is observed between peaks A \& B and marked by SD. The ratio of peak A to peak B is $\sim$0.9.       
  
  The CS using DFT are shown in Fig. 3(b). Firstly for VB, it is observed from the figure that the positions of two peaks are fairly good matched with the peak A and peak B within the experimental accuracy \cite{ley}. But, the dip (marked by arrow) in CS is dipper than ES. Similarly for CB, one hump and one peak are obtained at almost same position of hump C and peak D. One dip (marked by arrow) is observed in CS, which is masked in ES. One can also see the monotonically increased behaviour above $\sim$9.0 eV. Thus, DFT provides good explanation of ES except the line shape between peaks A \& B and hump C \& peak D. However, it is known that the consideration of electron-electron interaction in DFT for 3$d$ electrons is not adequate. This may be the reason for getting improper estimation of the depth of dip in CS. At this stage, it is important to note that we have also calculated the spectra using Perdew-Burke-Ernzerhof (PBE)\cite{perdew} which is one of the most used generalized gradient approximation (GGA) based functional. This CS is found to be almost the same as that obtained from LDA. Hence, it would be interesting to see whether a better technique like DFT+$U$ will improve the shallowness of dips. 
 
  In DFT+$U$, generally $U_{diag}$ is used for the value of $U$ parameter \cite{jepsen}. Therefore, the $U$ value of 3.4 eV is used for DFT+$U$, which is the value of $U_{diag}$. The CS (black solid line) thus obtained are shown in Fig. 3(c). It is observed from the figure that the one peak and one hump of CS correspond to peak B and hump C are good matching with ES. But in VB, the peak corresponds to peak A and the dip is totally washed out. For CB, the peak corresponding to peak D is shifted by $\sim$0.9 eV towards the higher energy. Therefore, this CS are no longer in good agreement with ES suggesting that $U_{diag}$ is quite large for DFT+$U$ calculation. This observation may not be surprising, as DFT+$U$ is a static mean field theory, where the screening among the $d$-electrons is not explicitly considered. So, any value of $U$, which is obtained by excluding this screening, is expected to be in the higher side. In this case, only that value of $U$ will be appropriate, which is calculated by considering the screening from all electron. Therefore, the value of $W$ (= 1.1 eV) may be the appropriate for $U$ parameter in DFT+$U$ calculation. Hence, the $U$ value of 1.1 eV is used for obtaining the CS (red dashed line), which are shown in Fig. 3(c). All the peaks' position, peaks' height and the hump's position of this CS are same as DFT, except the peak corresponds to peak D is slightly shifted towards the higher energy. In CB, the monotonically increased CS are shifted far away from the dashed line E. However, the shallowness of dip (marked by arrow) in VB is slightly improved in comparison to DFT. This method also fails to describe the shallowness of dips properly. 

  At this stage, it is important to note that the electron-electron interaction in DFT and DFT+$U$ depends only on crystal momentum \textbf{k}. It is well known that the many-body interaction among electrons is normally a function of \textbf{k} and $\omega$. Therefore, any method takes care of this aspects of interaction may provide correct electronic structure of material. However, this aspect of the interacting electrons is implemented in GWA using many-body perturbation theory. Here, one should keep in mind that the spectra obtained from $GW$ based calculations are nothing but the DFT DOS corrected by $GW$ energies. Thus, at first $G_0W_0$, which is the simplest $GW$ based method, is used for obtaining the CS. This CS are plotted in Fig. 3(d). It is observed from the figure that $G_0W_0$ gives similar features as obtained from DFT, except after 8.0 eV. After 8.0 eV, the monotonically increasing behaviour becomes more prominent. Also, this is shifted by $\sim$0.7 eV towards the higher energy as compare to the dashed line E. But, the shallowness of dips (marked by arrows) have not properly estimated. At this point, one should keep in mind that $G_0W_0$ method does not provide the proper many-body excitation $e.g.$, satellite feature/incoherent states, which may be the reason for not providing proper depth of the dips. Since, $GW_0$ method includes some of these aspects, where $G$ is updated self-consistently for some fixed $W_0$. Therefore, this is used for obtaining the CS, which are shown in Fig. 3(e). It is found from the figure that one peak and one hump corresponding to peak B and hump C, respectively, are in fairly good agreement with ES. After 0.7 eV, the increased CS (marked by arrow) are nicely matched with ES. However, The dip (marked by arrow) in VB is almost vanished due to the overestimation of states in this region. It is also observed that the other two peaks' positions, which are correspond to peak A and peak D, are showing very much off from the ES. Thus, this method badly fails in representing the ES. It is evident from above discussion that all these methods are not adequate for proper explanation of ES. Therefore, the DFT+DMFT, which is one of the most advanced electronic structure technique, may be needed for better understanding of ES.    

\begin{figure}
  \begin{center}
    \includegraphics[width=0.75\linewidth, height=4.5cm]{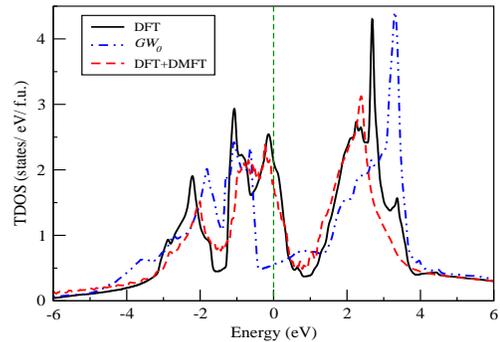} 
    \caption{(Colour online) Total density of states using DFT (black solid line), $GW_0$ (blue dotted dash line) and DFT+DMFT (red dashed line) at 300 K. Zero energy represents the Fermi level.}
    \label{fig:}
  \end{center}
\end{figure}

\begin{figure}
  \begin{center}
    \includegraphics[width=0.9\linewidth, height=6.5cm]{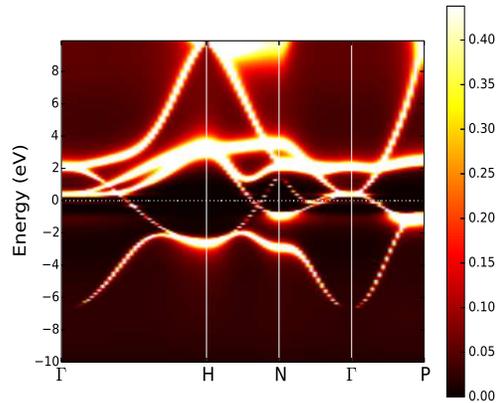} 
    \caption{(Colour online) Momentum-resolved many-body spectral function at 300 K. Zero energy represents the Fermi level.}
    \label{fig:}
  \end{center}
\end{figure}

   The DFT+DMFT takes care of the $d-d$ electrons screening. Thus in this case, $U_{diag}$ (= 3.4 eV) may be needed for the proper value of $U$ parameter. So, the $U$=3.4 eV is used in DFT+DMFT for computing the CS, which are shown in Fig. 3(f). It is observed from the figure that all the features including the monotonically increasing behaviour after 8.0 eV are nicely matched with the ES within the experimental uncertainty of $\sim$0.1 eV \cite{ley,speier}. In order to compare both the line shape and shallowness of dips within the energy $\sim$ -4.0 eV to $\sim$4.0 eV, three CS obtained from DFT, DFT+$U$ and DFT+DMFT are plotted together in the inset of Fig. 3(f). It is clear from the inset that in case of DFT+DMFT, the states are less populated $\sim$ -1.0 eV and $\sim$ -2.5 eV, whereas the states are more populated $\sim$ -1.5 eV as compare to DFT and DFT+$U$. Similarly, it is observed from the inset of CB that the states are more (less) populated from $\sim$0.5 ($\sim$2.5) eV to $\sim$2.0 ($\sim$3.5) eV. Thus, the spectral weight transfer from more populated to the less populated region appears to be responsible for proper estimation of shallowness of the dips along with overall improvement of the line shape of CS.

\begin{figure}
  \begin{center}
     \includegraphics[width=0.75\linewidth, height=4.5cm]{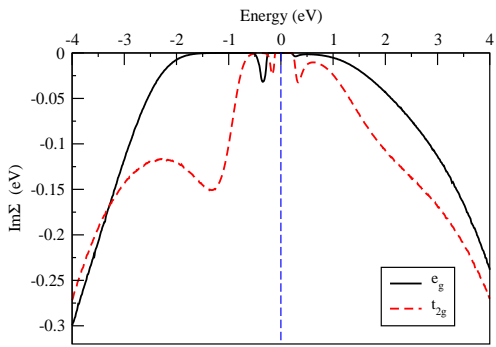} 
     \caption{(Colour online) Imaginary parts of self-energy ($Im\, \Sigma (\omega)$) as function of energy at 300 K for $e_g$ (black solid line) and $t_{2g}$ (red dashed line) orbitals. Zero energy represents the Fermi level.}
    \label{fig:}
  \end{center}
\end{figure}

  Now, in order to explore the electronic states of V in more detail, three TDOS calculated by DFT, $GW_0$ and DFT+DMFT are plotted within the energy -6.0 eV to 6.0 eV in Fig. 4. Here, it is important to note that the DFT+DMFT gives the information about many-body states with finite lifetime and represented by spectral function (A($\textbf{k},\, \omega$)), whereas the DOS refers to one-particle states with infinite lifetime. To distinguish A($\textbf{k},\, \omega$) with ${\sum\limits_{\textbf{k}} }$ A($\textbf{k},\, \omega$), normally different terminology is used. For example in the review article of Imada $et$ $al$ \cite{mimada} and in the paper of Haule $et$ $al$ \cite{yee} the terminology \textquoteleft DOS\textquoteright \, is used for ${\sum\limits_{\textbf{k}} }$ A($\textbf{k},\, \omega$). Here, we have followed this convention. It is observed from the figure that around -1.5 eV, $GW_0$ (DFT) populates more (less) states than DFT+DMFT, where intermediate amount of states are achieved. The similar behaviour is also found around 1.0 eV region. Thus in overall, $GW_0$ (DFT) overestimates (underestimates) the result for obtaining the TDOS, while DFT+DMFT properly estimates the states in the energy region of the dips. In order to understand the nature of these states in region $\sim$ -1.5 ($\sim$1.0) eV of VB (CB), the momentum-resolved spectral function calculated using DFT+DMFT along the high symmetric k-directions is shown in Fig. 5. To make the better insight of this figure, we need to understand A($\textbf{k},\, \omega$), which is defined as,
\begin{equation}
A{(\mathbf{k},w)} ={\Huge  \frac{1}{\pi} \frac{\mid Im\, \Sigma (\omega) \mid}{[\omega - \varepsilon^0_\mathbf{k} - Re\, \Sigma (\omega)]^2 + [Im\, \Sigma (\omega)]^2}}
\end{equation}  
where $\omega$ is real frequency, $\varepsilon^0_\mathbf{k}$ is the energy of a single non-interacting electron with crystal momentum $\mathbf{k}$ and $Im\, \Sigma (\omega)$ ($Re\, \Sigma (\omega)$) is imaginary (real) part of $\Sigma$($\omega$). Here, the pole ($i.e.$ $\omega = \varepsilon^0_\mathbf{k} + Re\, \Sigma (\omega)$) gives the energy position of the quasiparticle states. Thus in DFT+DMFT, energy position of the quasiparticle states is expected to change from their DFT energy position. This may lead to change in the population of states in the given energy window. The coherent and incoherent states may be defined in terms of the lifetime of the quasiparticle excitations depending on $Im\, \Sigma (\omega)$. When, $Im\, \Sigma (\omega)$ is quite small (large along with $\omega \neq \varepsilon^0_\mathbf{k} + Re\, \Sigma (\omega)$) then corresponding excitations will be coherent (incoherent). The presence of incoherent states in the energy from $\sim$ -2.0 ($\sim$0.5) eV to $\sim$ -1.0 ($\sim$1.5) eV around N- and P-points ($\Gamma$-point) can be clearly seen from Fig. 5. Hence, the incoherent states are more populated in these energy window resulting in improvement of peaks' shallowness in the CS. The presence of incoherent states in this energy window can also be observed from Fig. 6, where $Im\, \Sigma (\omega)$ is plotted as a function of energy for $e_g$ and $t_{2g}$ orbitals within the energy window -4.0 eV to 4.0 eV. Here, $Im\, \Sigma (\omega)$ shows negligibly small value for the $e_g$ orbitals from $\sim$ -2.0 eV to $\sim$1.0 eV suggesting coherent nature of $e_g$ orbitals. However, the value of $Im\, \Sigma (\omega)$ for $t_{2g}$ orbitals is $\sim$ -110.0 meV for this energy window, whereas it is lying from $\sim$ -10.0 meV to $\sim$ -70.0 meV for the energy window of $\sim$0.5 eV to $\sim$1.5 eV. Such a large value of $Im\, \Sigma (\omega)$ (specially for VB) is suggesting the incoherent part of $t_{2g}$ states in these energy window. Finally, we want to present $m^*$, which is defined as,
\begin{eqnarray} 
\frac{m^*}{m_{DFT}} = 1 - \frac{dRe\, \Sigma (\omega)}{d\omega}\mid_{\omega=0}
\end{eqnarray}    
The calculated values of $\frac{m^*}{m_{DFT}}$ using this relation for $e_g$ and $t_{2g}$ orbitals are found to be $\sim$1.14 and $\sim$1.22, respectively.

\section{Conclusions} 

  In the present work, a detailed electronic structure calculations for V have been performed using DFT, DFT+$U$, $G_0W_0$, $GW_0$ and DFT+DMFT techniques. The values of $W$, $U$ and $J$ are calculated by cRPA method, which are found to be $\sim$1.1 eV, $3.4$ eV and $\sim$0.52 eV. The large $\omega$ dependence of $U$ and $W$ are suggesting V as a correlated electron system. The calculated spectra obtained from different $ab$ $initio$ methods are compared with the experimental spectra (ES). The value of $W$ ($U$) is found to be suitable for DFT+$U$ (DFT+DMFT). All the technique except $GW_0$ provides the good estimation of the peaks' positions. However, they fail to provide proper shallowness of the dips as observed in ES, except for DFT+DMFT. The incoherent $t_{2g}$ states play important role in improving this shallowness of the dips. This result suggest the importance of advance technique like DFT+DMFT in proper understanding of the occupied and unoccupied electronics states for one of the simplest material.  

\section{Author contribution statement}

This problem is solely formulated by S.K.P. Under his guidance, all the numerical calculations, careful study of data and preparing the manuscript are done by A.S. After the discussion between both of A.S. and S.K.P, the results and the comments on manuscript at each stages of the revision are prepared.

\end{document}